\documentclass[useAMS,usenatbib]{mn2e}
\usepackage{graphicx}

\title[Ultimate age-dating for galaxy groups]{Ultimate age-dating method for galaxy groups; clues from the Millennium Simulations}
\author[M. Raouf et al.]{Mojtaba Raouf $^{1}$\thanks{E-mail:
m.raouf@ipm.ir},
Habib G. Khosroshahi$^{1}$, 
Trevor J. Ponman$^{3}$,
Ali A. Dariush$^{2}$,
\newauthor
Alireza Molaeinezhad $^{1}$, 
Saeed Tavasoli$^{1}$
\\
$^{1}$School of Astronomy, Institute for Research in Fundamental Sciences (IPM), Tehran, 19395-5746, Iran\\
$^{2}$Institute of Astronomy, University of Cambridge, Madingley Road, Cambridge CB3 0HA, UK\\
$^{3}$School of Physics and Astronomy, University of Birmingham, Birmingham, B152TT, UK
}

\begin{document}
\date{Accepted 2014 May 10.  Received 2014 May 10; in original form 2014 January 15}
\pagerange{\pageref{firstpage}--\pageref{lastpage}} \pubyear{2013}
\maketitle
\label{firstpage}

\begin{abstract}
There have been a number of studies dedicated to identification of fossil galaxy groups, 
arguably groups with a relatively old formation epoch. Most of such studies identify fossil 
groups, primarily based on a large luminosity gap, which is the magnitude gap between 
the two most luminous galaxies in the group. Studies of these types of groups in the 
millennium cosmological simulations show that, although they have accumulated a significant 
fraction of their mass, relatively earlier than groups with a small luminosity gap, this parameter 
alone is not highly efficient in fully discriminating between the "old" and "young" galaxy groups, 
a label assigned based on halo mass accumulation history. 

We study  galaxies drawn from the semi-analytic models of \cite{Guo11}, based on the 
Millennium Simulation. We establish a set of four observationally measurable parameters 
which can be used in combination, to identify a subset of galaxy groups which are old, with a very
high probability. We thus argue that a sample of fossil groups selected based on luminosity gap will result 
in a contaminated sample of old galaxy groups. By adding constraints on the luminosity of the brightest galaxy, and its offset from the group luminosity centroid,
we can considerably improve the age-dating.

\end{abstract}
\section{Introduction}

The age determination for galaxy systems in the hierarchical structure formation is not trivial because, 
in this paradigm, more massive galaxy systems such as galaxy clusters are formed through the 
mergers of smaller galaxy systems and are thus generally young galaxy systems. A galaxy group, 
however, can be recently formed or forming while some could be relatively old if they have not 
being subject to a substantial merging with other galaxy systems. In fact the answer to the question 
of age determination should come from the the cosmological simulations where evolutionary 
history of galaxy systems can be studied. The cosmological dark matter simulations and the 
implemented semi-analytic galaxy models offer tools, needed to develop an insight into this subject. 

\begin{figure*}
  \centering
\includegraphics[width=0.7\textwidth]{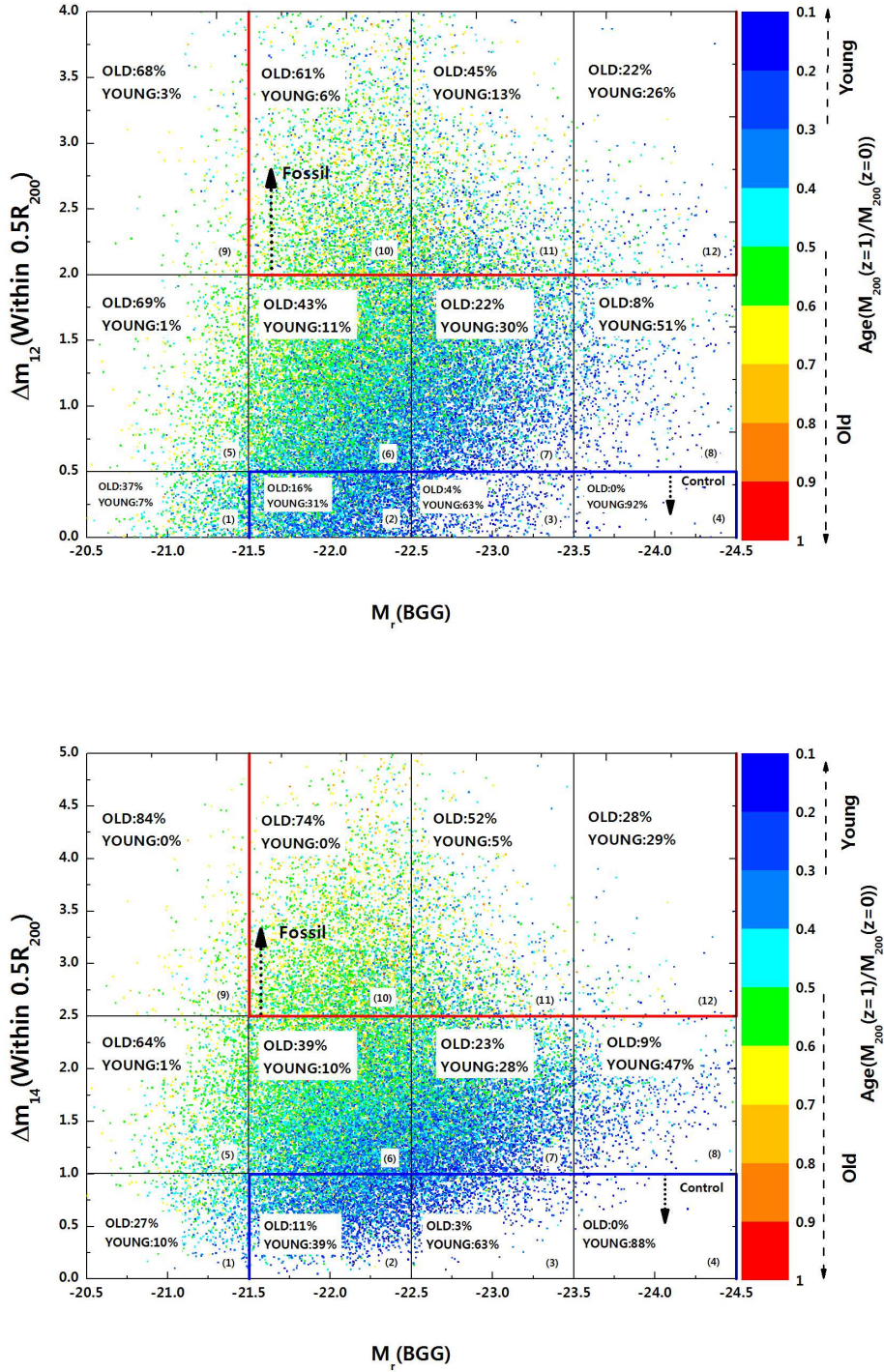}
\caption{Top: Distribution of the galaxy groups in the plane of luminosity gap $\Delta$m$_{12}$ 
within 0.5R$_{200}$ and the r-band magnitude of the Brightest Group Galaxy, $M_r(BGG)$, in 
the Millennium simulations with \citet{Guo11} semi-analytic model. Data point are colour-coded 
according to the ratio of the group halo mass at  redshift $z\approx1$ to its mass at z=0 ($\alpha_{0,1}$). 
The red box defines fossil groups region, e.g. groups dominated by a giant galaxy and 
$\Delta$m$_{12}\geq2$,  while the blue box defines control groups with $\Delta$m$_{12}\leq0.5$). 
The plane has been sub-divided into blocks within which the probability that the halo is old or 
young, is given. In this diagram panels (5), (9) and (10) contain mostly old systems while the 
panels (3), (4) and(8) are mostly occupied by young systems. By our definition, a group is old 
if its halo has over 50 per cent of its final mass at z=1 and its young if this fraction is less than 
30 per cent. Bottom: Same as the top panel, for the exception of $\Delta$m$_{14}$ within 
0.5R$_{200}$, is used (see the text), as an age indicator \citep{Dariush2010}. In this diagram 
panels (5), (9) and (10) contain mostly old systems while the panels (3), (4) and (8) are 
mostly occupied by young systems.}
\label{Deltam12_14}
\end{figure*}

Before these robust simulations became available in cosmological scales, an observational method for identification of old galaxy groups was developed in a pioneering study by \cite{Ponman1994}. This was motivated by earlier numerical simulations suggesting that in most cases, members of compact galaxy groups  could merge to form a single elliptical galaxy in a few billion years \citep{Barnes1989}. An elliptical galaxy formed  by the merger of such a group retains its X-ray  emitting halo  of hot gas, which is unaffected by  merging \citep{Ponman1994}. Such groups were called fossil groups in which the essential observational tracer has been identified as the large luminosity gap between the two brightest group members. 

According to the convention introduced by \citet{Jones2003} fossil groups have a magnitude gap of at least 2 magnitude within 0.5 virial radius and  $L_{X,bol}\approx 10^{42} $~h$_{50}^{-2}$ ~erg~s$^{-1}$. Since then there have been many studies focused on the detailed characterisation and properties of fossil groups \citep{Khosroshahi2004a,Sun2004,Ulmer2005,Khosroshahi2006}, based on X-ray and optical observations. However, due to limited number of adequate X-ray surveys and their suitability, most studies selected the magnitude gap as the fossil identifier in theoretical and observational studies \citep{Yoshioka2004,Milosavljevic2006,VandenBosch2007},  hydrodynamic simulations \citep{DOnghia2005}, dark matter simulations \citep{VonBenda-Beckmann2008,Cui2011} and combined with semi-analytical models for galaxies \citep{Sales2007,Diaz2008,Dariush2010}.
\citet{Khosroshahi2007} show that for a given optical luminosity, fossil groups are not only more X-ray luminous than the general population of galaxy groups, but also they have a more concentrated halo as well as hotter IGM for a given halo mass. The study of scaling laws of fossil groups also show that they mostly follow the trend of galaxy clusters which is likely to be driven by their dynamical relaxed state, although \citet{Voevodkin2010} have shown that in the cluster regime there are no noticeable difference between the X-ray luminosity of the fossils and non-fossils for a given optical luminosity. In addition, some other studies support the same conclusion employing other cluster samples \citep{Aguerri2011,Harrison2012,Girardi2014}.

One of the largest cosmological simulations, the Millennium Simulation \citep{Springel2005} joined 
with Semi-Analytical Models (SAMs) for galaxy formation, provide a useful tool to address open questions, regarding the age determination of the galaxy systems. \citet{Dariush2007} show that the luminosity gap is a fairly good indicator of the halo mass assembly, such that in galaxy groups with a large luminosity gap ($\Delta m_{12}\ge2$), the halos accumulate 50 per cent of their mass at current epoch by $z=1$ and thus are relatively older than their counterparts with small luminosity gap. \citet{Dariush2010} introduce an alternative optical criterion $\Delta$m$_{14} \geq 2.5$, i.e. the luminosity gap between the first and fourth brightest galaxies within $0.5R_{200}$, which is found to be more efficient in identifying early-formed groups than the conventional criterion, $\Delta$m$_{12} \geq 2.0$. 

Following the discovery of the fossils groups with different masses, from surveys or serendipitous observations, there has been attempts to study and compare various halo, IGM and galaxy properties of fossil and non-fossil groups \citep{Khosroshahi2007,Dupke2010,Miller2012}. The velocity dispersion of the member galaxies in groups \citep{Herbst2012,Madrid2013}, the globular cluster distribution and colour \citep{Alamo2012}, the galaxy luminosity function \citep{Cypriano2006,Lieder2013}, the AGN activities of the groups \citep{Hess2012,Miraghaee2013}, merger history \citep{Eigenthaler2013,Diaz2008} and halo concentration \citep{Khosroshahi2007,Deason2013}  are amongst these properties. The basic argument upon which these studies are based, is the early formation of fossils groups. The older age of fossil group has been argued in the earlier studies \citep{Ponman1994,Jones2003} and demonstrated in the cosmological simulations \citep{DOnghia2005,Dariush2007}.  However there is still room to develop age dating algorithms with a high efficiency using measurable parameters accessible from routine optical observations and other convenient methods.

In this paper, we first show the extend at which the large luminosity gap can discriminate old and the young galaxy groups. We then introduce a some parameters in which the old and young galaxy groups can be identified more distinctively. We finally define a parameter space which consists of optically measurable parameters of group galaxies, that allow age-dating with a very high efficiency. 

Section 2 describes Millennium Simulation and Semi-Analytical Model.  In section 3 we describe our multi-parameter analysis to search for old and young groups using the optical measurements. Finally in section 4 present summery of our result and given our conclusion. 

\section{The simulation}

In this study we use the public release of the Millennium Simulation(Hereafter: MS) \citep{Springel2005}. The cosmological model adopted in  simulation is a $\Lambda$CDM with the following parameters: ${\Omega}_m=0.25$, ${\Omega}_b=0.045$, ${\Omega}_{\Lambda}=0.75$, h=0.73, n=1,${\sigma}_8=0.9$ (note that the value of ${\sigma}_8$ is assumed to be greater than its present value of 0.82 given by WMAP9 that is not strongly affects in this study). The simulation box $(500 h^{-1} Mpc)^{3}$ contains $2160^{3}$ particles and presents the mass resolution of 8.6${\times}$ 10$^8$ ~h$^{-1}$~M$_{\odot}$.  

\begin{figure*}
\centering
\includegraphics[width=0.95\textwidth]{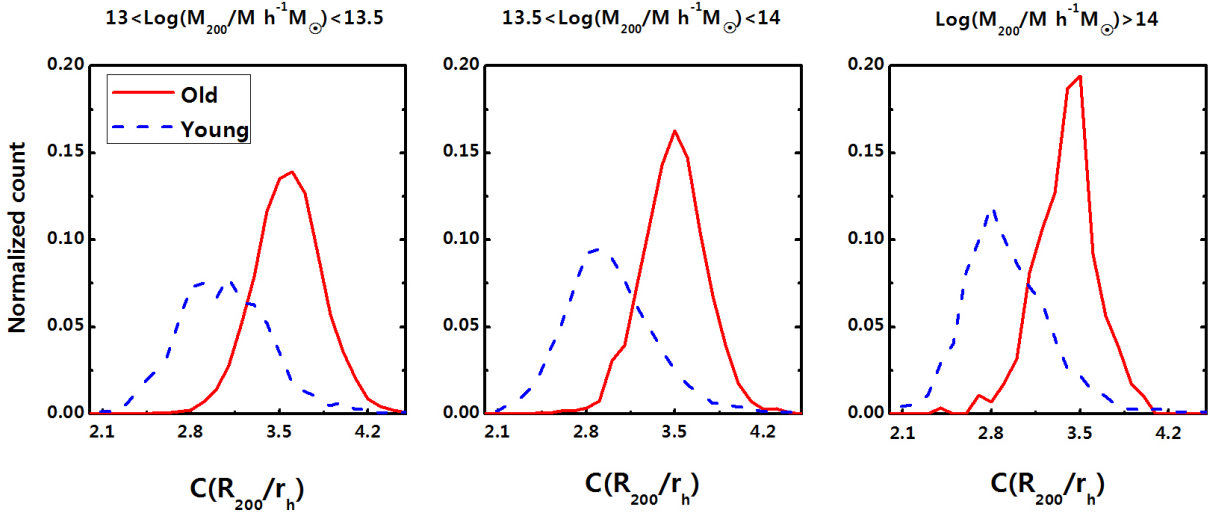}
\caption{A comparison between the halo concentration(C) of old (red-line) and young (blue-dash-line) groups in MS using \citet{Guo11} SAM for different halo masses. We estimate the halo concentration as the ratio between the $R_{200}$ and half-mass radius for all candidates for old and young systems. The old systems show higher halo concentration than the young galaxy systems.
}
\label{c_fig}
\end{figure*}
The dark matter merger trees within each simulation snapshot (64 snapshots) are expanded approximately logarithmically in time between $z=127$ and $z=0$ and extracted from the simulation using a combination of friends-of-friends (FoF) \citep{Davis1985} and SUBFIND \citep{Springel2001} halo finders algorithm. The gas and stellar components of galaxies in dark matter halos are constructed semi-analytically, based on differ in the phenomenological recipes. 

For galaxy properties, we use \citet{Guo11} semi-analytical model, in which
the treatments of many of the physical processes have been improved in 
comparison to an earlier model by \cite{DeLucia2007}. This model provides good fits to 
the observed luminosity and stellar mass functions of galaxies, from  the SDSS 
data and recent determinations of the abundance of satellite galaxies around 
the Milky Way and the clustering properties of galaxies as a function of stellar 
mass, as well. Data available in the Millennium database \citep{Lemson2006}, contains 
$\sim 51000$ halos with  masses above 10$^{13}$ ~h$^{-1}$~M$_{\odot}$  and  
$\sim5$ million galaxies from which we only select galaxies brighter than $-14$ in r-band 
absolute magnitude for completeness.

\section{Analysis and results}

\subsection{Age-dating based on luminosity gap}
\subsubsection{$\Delta$m$_{12}$}
As mentioned earlier, the magnitude gap between the brightest and second brightest galaxies in a galaxy group, is often used as an indicator of the dynamical age of groups of galaxy. \citet{Dariush2010} investigated the assembly of groups and clusters of galaxies using the Millennium dark matter simulation and semi-analytic catalogues of galaxies. The study aimed at verifying the argument that galaxy groups with a large magnitude gap are formed earlier than the galaxy groups with a small magnitude gap. They used mass accumulation of the group halo as a proxy for early/late halo formation. They selected galaxy groups and clusters at the present time with dark matter halo mass $M(R_{200}) > 10^{13}h^{-1}M_{\odot}$, and trace their properties to $z\simeq1$.  In addition they applied an X-ray luminosity criteria to keep those galaxy systems for which the X-ray luminosity $L_{X,bol} > 0.25 \times 10^{42}h^{-2} erg s^{-1}$ at redshift $z=0$. They argued that while it is true that a large magnitude gap between the two brightest galaxies of a particular group often indicates that a large fraction of its mass was assembled at an early epoch, it is not a necessary condition. More than 90\% of fossil groups defined on the basis of their magnitude gaps (at any epoch between $0 < z < 1$ ) cease to be fossils within 4 Gyr, mostly because other massive galaxies are assembled within their cores, even though most of the mass in their haloes might have been assembled at early times.\\

As a definition here \citep[and][]{Dariush2007}, a galaxy group which formed more than 50 per cent of its total mass by $z\approx1$ is labled as old. A group is labelled young if less than 30 per cent of its final mass is formed by $z\approx1$. In Figure \ref{Deltam12_14} top panel, we present the luminosity gap $\Delta$m$_{12}$ (within 0.5R$_{200}$) as a function of the Brightest Group Galaxy (BGG) magnitude in r-band, $M_r(BGG)$, given for all 39132 groups (i.e. groups with $M(R_{200}) \geq $ 10$^{13}$~h$^{-1}$M$_\odot$ and exist within both $z=0$ and $z=1$ ) using \citet{Guo11} SAM at the present epoch$(z=0)$. The groups are colour-coded according to their $\alpha_{0,1}$ parameters(defined as "age" of systems), where $\alpha_{0,1} = M_{z\approx1}/M_{z=0}$. 
The horizontal line subdivides groups into magnitude gap bins. Those with $\Delta$m$_{12}\geq2$ are conventional fossil groups. Groups with $\Delta$m$_{12}\leq0.5$ are labelled as control groups which known to be young galaxy groups \citep{Dariush2007,Dariush2010}. Vertical lines bin the groups according to the luminosity of their brightest galaxy (BGG) in 4 magnitude bins from -20.5 to -24.5. We note that the systems with large magnitude gap and faint BGGs are modest galaxies with some dwarf satellites, similar to the Milky Way. As these systems do not present galaxy groups, satisfying fossil groups condition, we limit our analysis to BGGs which are at least as bright as $M_{R} < -21.5$, i.e giant galaxies. This diagram shows that the galaxy luminosity gap combined with the luminosity of the brightest group galaxy for all halo mass over $10^{13} M_{\odot} h^{-1}$ is success to identify old groups with a probability of  61 per cent  and  young galaxy group with a probability of 92 per cent.

\subsubsection{$\Delta$m$_{14}$}
 
The assembly time of a dark matter halo defined as the look-back time when its main progenitor reaches a mass that is more than 50 per cent of the present halo mass. \citet{Dariush2010} showed that the alternative optical criterion $\Delta$m$_{14} \geq 2.5$ (in the R-band, within $0.5R_{200}$ of the centre of the group) is more efficient in identifying early-formed groups than the conventional criterion $\Delta$m$_{12} \geq 2.0$ and it is more success to identify at least 50\% more early-formed groups. However, the conventional criterion ($\Delta$m$_{12} \geq 2.0$) performs marginally better at finding early-formed groups at the high-mass end of groups. Figure \ref{Deltam12_14} Bottom panel shows distribution of luminosity gap $\Delta$m$_{14}$ within $0.5R_{200}$ combined with the Brightest Galaxy Group (BGG) magnitude in r-band. In this diagram panels (5), (9) and (10) contain mostly old systems while the panels (3), (4) and (8) are mostly occupied by young systems. The mentioned panels can identify old groups with a probability of at most 74 per cent  and  young galaxy group with a probability, up to 88 per cent.
\begin{figure*}
\centering
\includegraphics[width=0.95\textwidth]{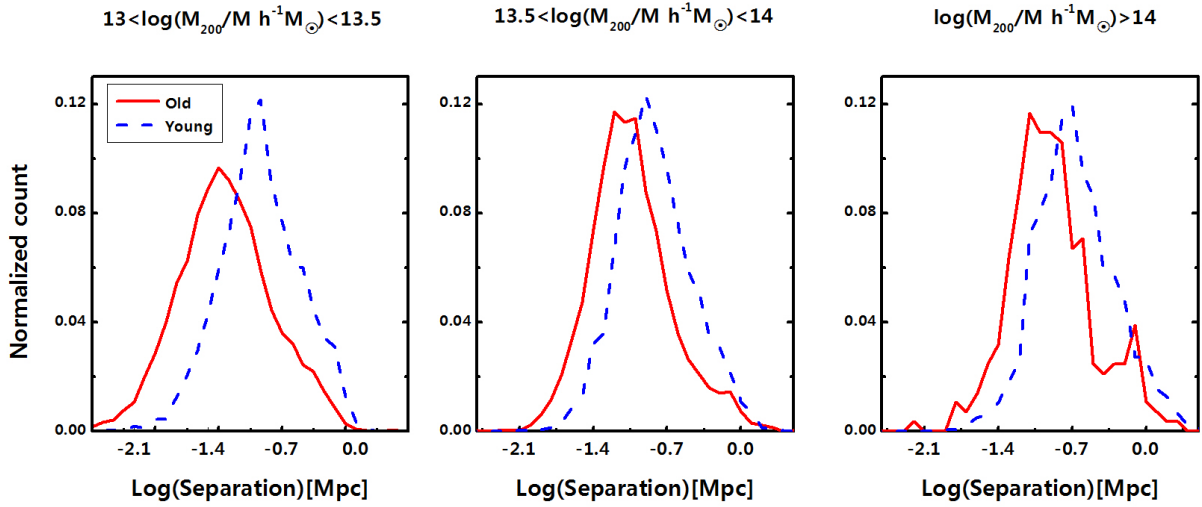}
\caption{ The distribution of luminosity de-centring, the distance between the location of the BGG and the luminosity centroid (luminosity de-centring) in various mass bins. Galaxy systems with a large luminosity de-centring are dynamically unrelaxed and thus younger.
}
\label{dec_fig}
\end{figure*}
\subsection{Other age indicators}

\subsubsection{Halo concentration}

 It has been shown that the halo concentration is linked to the epoch of the halo formation \citep{Wechsler2002}. There are different ways to estimate halo concentration in the literature. A popular method is by fitting a NFW \citep{NFW96} profile shows in eq. \ref{NFW_profile} to the halo density distribution
\begin{equation}
\rho_{NFW}=\frac{\rho_s}{x(1+x)^{2}}
\label{NFW_profile}
\end{equation}
 
where $x\equiv r/r_{s}$ and $\rho_s$ and $r_{s}$ are the characteristic density and radius. Then the halo concentration($C_{200}$) is found as eq.\ref{concentration}
\begin{equation}
 C_{200} = \frac{R_{200}}{r_{s}}
 \label{concentration}
\end{equation}
where $C_{200}$ is the ratio of $R_{200}$ as enclosing a mean density of 200 times the critical rather than the cosmological mean density to the characteristic radius $r_{s}$.

Considering the relation between halo mass and concentration, lower mass halos universally form at earlier epochs relative to higher mass halos \citep{Gao2008,Prada2012}. Therefore, the lower mass halos must be more concentrated \citep{Wechsler2002}. Due to this close connection between halo concentration and assembly history; group halos that formed most of their halo mass at early times are also more concentrated in comparison to the halos with later formation epoch \citep{Deason2013}. Observationally, for a given optical luminosity, fossil groups have  more concentrated halos as well as hotter IGM for a given halo mass \citep{Khosroshahi2007}.

Using the Millennium simulations, \citet{Ludlow2012} demonstrated that, the virial-to-half mass radius, $R_{200}/r_{h}$ in which  $r_h$ is the half-mass radius, is a reliable tracer for the halo concentration. We thus estimate the halo concentration as the ratio between the $R_{200}$ and half-mass radius for all candidates for old and young systems. As expected, in Figure \ref{c_fig}, the halos of early formed groups show higher concentration compared to the same in late formed groups. While the bimodal distribution of the halo concentration between the old and the young groups (Figure \ref{c_fig}) appears to be very attractive for discriminating these two type of groups,
there are significant observational limitations.

Observationally, the halo concentration can be obtained from any observations that can provide halo profile measurement, such as the the X-ray observations of the hot gas, strong/weak gravitational lensing and galaxy distribution within the group/cluster halo. 

The X-ray method of measuring the halo mass/density profile is built upon the hydrostatic equilibrium assumption where the gravitational contraction is balanced by the hot gas pressure. This requires measurements of the gas density and the temperature profile which can be obtained from X-ray observations. In practice, additional assumptions are made, such as the spherical symmetry \citep{Khosroshahi2004a, Humphrey2006} to obtain the halo mass/density profile \citep{Fabricant1984, Pratt2005,Khosroshahi2006a}. Usually a NFW profile (eq. \ref{NFW_profile}) is fitted to the total gravitational mass density which results in the halo concentration parameter according to eq. \ref{concentration}. 

The gravitational lensing method of mass profile reconstruction, is based on the observations of strong arc shaped features of a distant background source \citep{Fort1994} or the observation of lens induced shear on the shape of the source galaxies by a lens galaxy cluster  \citep{Miralda1991,Kaiser1993,Wilson2001}. These methods are observationally demanding, requiring high quality photometric and spectroscopic measurements, and can only provide a projected mass density for the lensing cluster  \citep[see][]{Leauthaud2007,Leauthaud2010}. In particular the case of galaxy groups is challenging due to a low lensing signal and stacking of groups may be necessary \citep{Hoe01}. The lensing method of halo mass measurement is only efficient in intermediate redshift range \citep{Smith2005,Oguri2010}, while there are many sources of systematic errors in weak lensing \citep{Mandelbaum2005, Johnston2007}. 

There are a limited number of observations which can provide halo concentration of galaxy groups and clusters with a reasonable uncertainty.  One of the largest samples of galaxy clusters, studied as part of the LoCuSS project \citep{Smith2010}, points at limited statistics on galaxy clusters for which the halo concentration is measured and also a typically  20 per cent uncertainty in X-ray measurements of the halo concentration, $C_{500}$ \citep{Sanderson2009}. Similarly, \cite{Khosroshahi2006a} in the study of fossil galaxy cluster reported a large uncertainty in the concentration, as high as $\sim 40 $ per cent. Thus,  halo concentration measurements obtained from gravitational lensing and/or X-ray observations may be subject to large uncertainties \citep{Mandelbaum2005,Mandelbaum2006a,Mandelbaum2006b,Johnston2007,Oguri2010,Shan2010a,Shan2010b}. 

\begin{figure}
\includegraphics[width=0.55\textwidth]{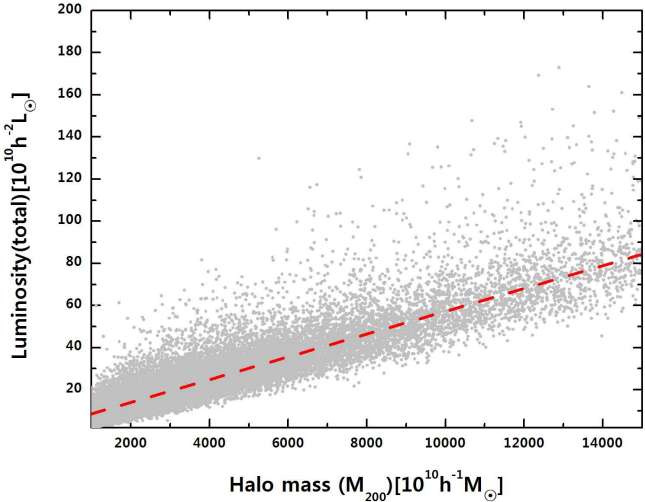}
\caption{
The relation between total optical luminosity within $R_{200}$  for each
semi-analytic group and its halo mass. The red dashed line, fitted to the
data, implies that masses of  $10^{13}h^{-1}M_{\odot}$, $10^{13.5}h^{-1}M_{\odot}$ correspond to total luminosities of $1.9\times 10^{10}h^{-2}L_{\odot}$, $20.27\times 10^{10}h^{-2}L_{\odot}$,  $57.2\times 10^{10}h^{-2}L_{\odot}$, respectively.}
\label{lum_mass}

\end{figure}
\begin{figure*}
\includegraphics[width=0.6\textwidth]{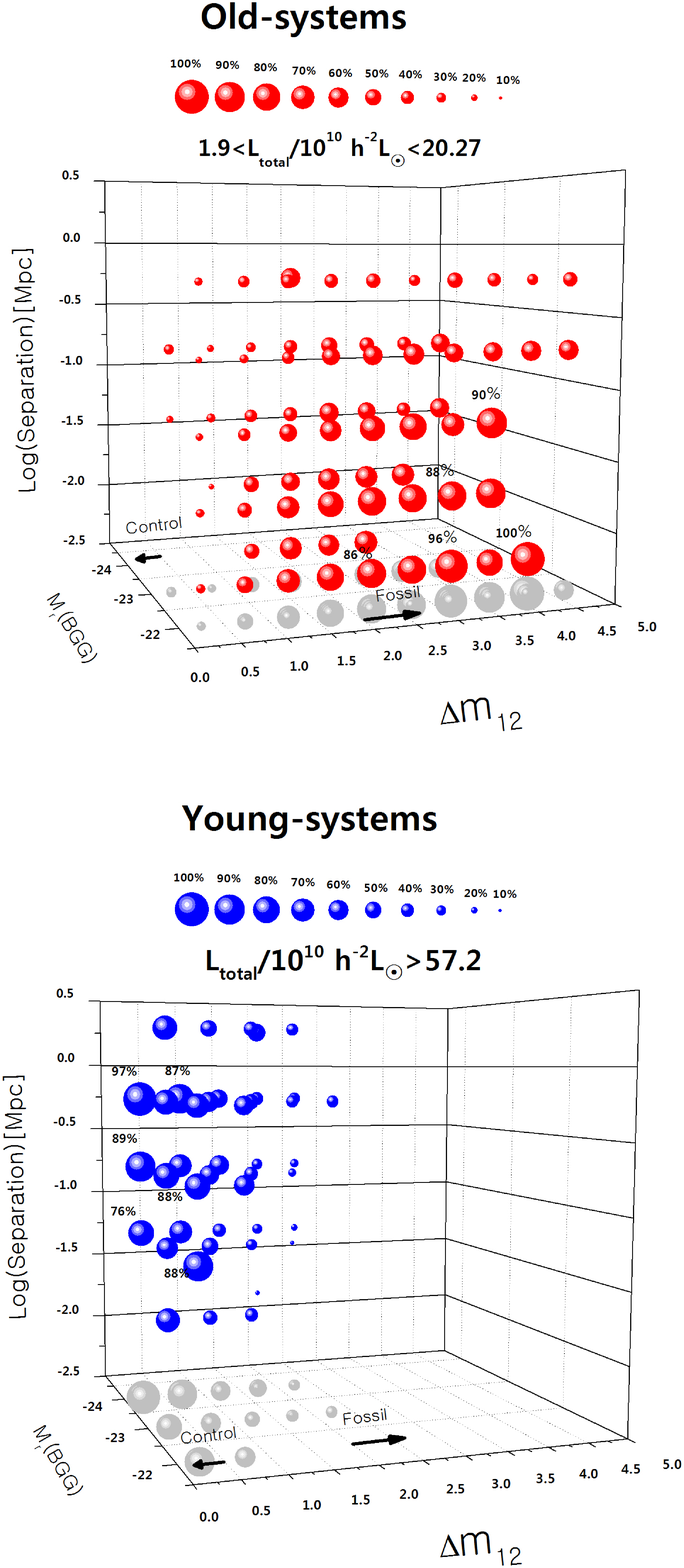}
\caption{The probability for a system to be considered as old (top) and young (bottom) based on the observable 4-dimensions parameter space. The parameter space consists of the total group optical luminosity, luminosity of the BGG, luminosity gap  and luminosity de-centring. Size of the symbols (balls) indicate the oldness probability for a given sub-region in the parameter space. For example within the total luminosity range 1.9$\times$10$^{10}$~h$^{-2}$L$_\odot$ through 20.27$\times$ 10$^{10}$~h$^{-2}$L$_\odot$, galaxy groups with Log(separation)($-2\ge Log(separation)\ge -2.5$) and M$_r$(BGG)$\approx -22$ mag, and $\Delta m_{12}\approx 4.5$ are 100 per cent old(see Table 2), according to our definition of age, based on mass accumulation. 
As seen low total luminosity systems show the highest population of old galaxy groups. In addition for a galaxy group to be old, the luminosity gap should be large while the luminosity segregation should be small. Same as top panel for young galaxy groups. Largest symbol indicates highest probability of galaxy groups been young. As shown youngest systems are found in high total luminosity galaxy groups where the luminosity gap is small and the luminosity segregation is large.}
\label{plot_old_young}
\end{figure*} 

\subsubsection{Luminosity de-centring}

Typically, galaxy formation models assume that the brightest group galaxy is located at the centre of group halos, however, \cite{Skibba2011} demonstrated that the BGG may not always be the central galaxy. In fact, merging systems are dynamically unrelaxed which can result in a significant separation between the brightest galaxy and the halo centre. This quantity has been used as a tracer of the dynamical age of the galaxy groups. For instance, \citet{Rasmussen2012} in a study of ongoing star formation of groups, use the separation of the brightest group galaxy from the luminosity centroid of member galaxies in the group as an age indicator for the galaxy groups. The X-ray emission peak, or the centroid, and the mass centroid from the gravitational lensing observations, have also been used as indicators of the halo centre \citep{Dietrich2012, Oguri2010}.

One of the interesting properties of the fossil groups is revealed by their X-ray morphology. The X-ray peak coincides very well with the centre of the dominant giant elliptical galaxy \citep{Khosroshahi2007}. Relaxed morphology of the clusters or absence of substructures are good indicators of the dynamical state of the galaxy systems \citep{Smith2005}. \citet{Smith2010} shows that clusters with dominant galaxy at the centre are the one with least substructure and are thus dynamically old.

In the absence of X-ray or lensing mass maps, one can resort to optical luminosity distribution. We thus define the luminosity weight of each groups as $X_{L}= \sum X_{i} L_{i}/ \sum L_{i}$, where $L_{i}$ is the luminosity of galaxy in group in r-band and $X_{i}$ is the projected coordinate of each galaxy. The de-centring parameter is thus defined as the projected physical separation between the luminosity centroid of the group and the location of the BGG. As shown in Figure \ref{dec_fig} younger systems show larger separation between BGG and the luminosity centroid of the galaxies in comparison to the old groups. This parameter can also be replaced with the separation between the BGG and other halo centroid indicators such as the one obtained from gravitational lensing and/or X-ray observations, however, our choice has the advantage of being optically measurable. 

\begin{table*}
\begin{center}
\centering
\caption{Selected regions in the 4-dimensional parameter space with highest probability of old and young groups: Column 1 gives the range of r-band total luminosity; Column 2 designates the separation between BGG and the Luminosity Centroid; Column 3 gives the magnitude gap between the two most luminous galaxies in the group; column 4 gives r-band absolute magnitude of the BGG, for the probability specified in column 5. Column 6 gives the mass to light ratio for the groups in the given region in the parameter space.} 

    \begin{tabular}{cccccc}
        \hline
         Group luminosity         & $log(Separation)$ & $\Delta$m$_{12}$      & $M_r(BGG)$   & probability(old/young) & $\frac{M_{200}}{L_{r}}$ \\ 
        $ [10^{10} h^{-2}L_{\odot}]  $ & [Mpc]  & [mag]   & [mag]   & [per cent]      & [Ratio$\pm\sigma]$ \\ 
         
         \hline\hline
         1.9$\le L < $20.27      & $-$2.5 -- $-$2.0        & 4.0 -- 4.5   & $-$21.5 --  $-$22.5     &  100\% old	 &  223.2$\pm$53.0   \\ 
         1.9$\le L < $20.27      & $-$2.5 -- $-$2.0         & 3.0 -- 3.5   & $-$21.5 --  $-$22.5     &  96\% old   & 217.2$\pm$45.0  \\ 
         1.9$\le L < $20.27      & $-$1.5 -- $-$1.0        & 3.5 -- 4.0   & $-$21.5 --  $-$22.5     &  90\% old   & 178$\pm$34.4  \\
         1.9$\le L < $20.27      & $-$2.0 -- $-$1.5         & 3.5 -- 4.0  & $-$21.5 --  $-$22.5     &  88\% old   & 222$\pm$57.4  \\
         1.9$\le L < $20.27      & $-$2.5 -- $-$2.0         & 2.0 -- 2.5   & $-$21.5 --  $-$22.5     &  86\% old   & 214.7$\pm$51.0  \\
         1.9$\le L < $20.27      & $-$2.0 -- $-$1.5         & 3.0 -- 3.5   & $-$21.5 --  $-$22.5     &  86\% old   & 219.3$\pm$51.0  \\
         1.9$\le L < $20.27      & $-$2.5 -- $-$2.0         & 2.5 -- 3.0   & $-$21.5 --  $-$22.5     &  85\% old   & 209$\pm$39.0  \\
         1.9$\le L < $20.27      & $-$2.0 -- $-$1.5         & 2.0 -- 2.5   & $-$21.5 --  $-$22.5     &  84\% old   & 206$\pm$49.0  \\
         1.9$\le L < $20.27      & $-$2.0 -- $-$1.5         & 2.5 -- 3.0   & $-$21.5 --  $-$22.5     &  83\% old   & 213$\pm$45.0  \\   
         1.9$\le L < $20.27      & $-$2.5 -- $-$2.0         & 1.5 -- 2.0   & $-$21.5 --  $-$22.5     &  80\% old   & 206$\pm$44.0  \\
         20.27$\le L < $57.2     & $-$1.5 -- $-$1.0         & 2.5 -- 3.0   & $-$22.5 --  $-$23.5     &  53\% old   & 181$\pm$49.8   \\ 
         20.27$\le L < $57.2     & $-$2.0 -- $-$1.5         & 2.0 -- 2.5   & $-$22.5 --  $-$23.5     &  50\% old   & 171.6$\pm$35.7   \\ 
         20.27$\le L < $57.2     & $-$1.5 -- $-$1.0         & 2.0 -- 2.5   & $-$22.5 --  $-$23.5     &  48\% old   & 176.8$\pm$35.0  \\ 
         20.27$\le L < $57.2     & $-$1.0 -- $-$0.5         & 2.5 -- 3.0  & $-$22.5 --  $-$23.5     &  47\% old   & 147$\pm$41.2 \\      
         20.27$\le L < $57.2     & $-$2.0 -- $-$1.5         & 1.5 -- 2.0   & $-$22.5 --  $-$23.5     &  43\% old   & 183.2$\pm$35.5  \\    
         20.27$\le L < $57.2     & $-$1.0 -- $-$0.5         & 0.0 -- 0.5   & $-$22.5 --  $-$23.5     &  58\% young & 151.3$\pm$ 35.0   \\ 
         20.27$\le L < $57.2     & $-$2.5 -- $-$2.0         & 0.0 -- 0.5   & $-$21.5 --  $-$22.5     &  56\% young & 176$\pm$30.0  \\ 
         20.27$\le L < $57.2     & $-$1.0 -- $-$0.5       & 0.0 -- 0.5   & $-$21.5 --  $-$22.5     &  54\% young & 150$\pm$35.3   \\ 
         20.27$\le L < $57.2     & $-$1.5 -- $-$1.0         & 0.0 -- 0.5   & $-$21.5 --  $-$22.5     &  51\% young & 162.4 $\pm$35.0 \\      
         20.27$\le L < $57.2     & $-$0.5 -- 0.0          & 3.0 -- 3.5   & $-$22.5 --  $-$23.5     &  48\% young & 108.5$\pm$43.7 \\         
         20.27$\le L < $57.2     & $-$0.5 -- 0.0          & 4.0 -- 4.5   & $-$21.5 --  $-$22.5     &  43\% young & 121.5 $\pm$34.0 \\                               
        $L > $57.2               & $-$0.5 -- 0.0          & 0.0 -- 0.5   & $-$23.5 --  $-$24.5     &  97\% young & 179$\pm$25.5    \\ 
        $ L > $57.2              & $-$1.0 -- $-$0.5         & 0.5 -- 1.0   & $-$23.5 --  $-$24.5     &  89\% young & 186.5$\pm$26.0 \\
        $ L > $57.2              & $-$1.5 -- $-$1.0         & 0.0 -- 0.5   & $-$21.5 --  $-$22.5     &  88\% young & 176.2$\pm$27.2 \\
        $ L > $57.2              & $-$0.5 -- 0.0          & 0.5 -- 1.0   & $-$23.5 --  $-$24.5    &  87\% young & 166.1$\pm$31.0 \\
        $ L > $57.2              & $-$1.0 -- $-$0.5         & 0.0 -- 0.5   & $-$21.5 --  $-$22.5     &  77\% young & 155.7$\pm$33.8  \\
        $ L > $57.2              & $-$1.5 -- $-$1.0         & 0.0 -- 0.5   & $-$23.5 --  $-$24.5     &  76\% young & 187$\pm$21.5 \\  
       \hline
    \end{tabular}   
    \end{center}
\label{tab2}
\end{table*}

\subsection{Age-dating based on optical observations; A 4-dimensional parameter space}
Age-dating base on halo concentration and other halo parameters method will rarely be possible to apply in practice, thus we look for a method using only parameters which are commonly available.

A 4-dimensional parameter space can be constructed based upon galaxy properties and using optical observations, only.  It has been shown that the total luminosity of a galaxy group is a fair indicator of the halo mass \citep{Tavasoli2012}. Figure \ref{lum_mass} shows the correlation between groups total luminosity  (within $R_{200}$) and halo mass. Using a linear fit to our data we find that the mass ranges of $10^{13}h^{-1}M_{\odot}$, $10^{13.5}h^{-1}M_{\odot}$ and $10^{14}h^{-1}M_{\odot}$ generally correspond to total luminosity ranges of $1.9\times 10^{10}h^{-2}L_{\odot}$, $20.27\times 10^{10}h^{-2}L_{\odot}$,  $57.2\times 10^{10}h^{-2}L_{\odot}$, respectively. While the halo concentration is found to be a good age indicator, in this 4-dimensional approach, we assume that only photometric observations of galaxy groups is available. These observations should be such that the two brightest groups members can be identified, for instance through spectroscopic observations. The total luminosity can be estimated through red-sequence fitting or similar methods which allow statistical memberships identifications. 

Figure \ref{plot_old_young} shows the probability for a system to be considered as old (top) and young (bottom) based on the observable 4-dimensional parameter space. The parameter space consists of the total group optical luminosity, luminosity of the BGG, luminosity gap ($\Delta m_{12}$)  and luminosity de-centring. In this Figure, size of the symbols (balls) indicate the oldness probability for given sub-region in the parameter space. The luminosity de-centring and the radius at which the magnitude gap, $\Delta m_{12}$, has been estimated are all projected. We note that the $\Delta m_{14}$ improves the statistical probability of old/young groups (Figure \ref{Deltam12_14}), however, we prefer to use $\Delta m_{12}$ due to relative ease of its measurement in spectroscopic surveys. For instance the groups for which the $\Delta m_{14}$ is measured in Sloan Digitized Sky Survey (DR10) accounts for 10\% of the groups for which the $\Delta m_{12}$ can be obtained. In addition, $\Delta m_{12}$ has been used widely in the majority of the studies to date. 

As an example galaxy groups with a total luminosity of 1.9$\times$10$^{10}$~h$^{-2}$L$_\odot$ through 20.27$\times$ 10$^{10}$~h$^{-2}$L$_\odot$, and $-2\ge log(separation)\ge -2.5$ and M$_r$(BGG)$\approx -22$ mag, with a $\Delta m_{12}\approx 4.5$ are 100 per cent old, according to our definition of age, which is based on mass accumulation history of the halo. As Figure \ref{plot_old_young} reveals, systems with low total luminosity show the highest population of old galaxy groups. Moreover for a galaxy group to be considered as old, the luminosity gap should be large while the luminosity segregation should be small. 

Bottom panel in Figure \ref{plot_old_young} shows the same probability for young galaxy groups. Largest symbol indicates highest probability of galaxy groups been young. As shown youngest systems are found in high total luminosity galaxy groups where the luminosity gap is small and the luminosity segregation is large.

In Table 2. we report few of the highest probability regions to find old and young galaxy groups; and the average mass-to-light ratio($M_{200}/L_{r}$) for a given region. Statistically, the old group halos have higher mass-to-light ratio relative to the young group halos.

\section{Conclusions}

Assigning age to galaxy systems in the Lambda-CDM hierarchical structure formation paradigm is not trivial, but essential, specially when one aims to study the connection between the evolution of the halos to the evolution of galaxies and the properties of the intergalactic medium. We quantify the probability of finding galaxy groups with a given halo mass accumulation history, namely old and young galaxy groups, in the parameter space of  the luminosity gap and the BGG luminosity for a given halo mass. We show that there is a limited success in identifying galaxy groups based on the luminosity gap as the statistical probability of finding old groups using this parameter is only $60\%$, in the Millennium simulations (based on \cite{Guo11} semi-analytic model for the galaxies). 

We define a 4-dimensional parameter space consisting, the optical luminosity of the brightest group galaxy, the optical luminosity difference between the two brightest galaxies in the groups (the luminosity gap), the total luminosity of the group (as a proxy to the group mass) and the physical separation between the luminosity centroid of the galaxy group and its brightest group galaxy. Using this multi-dimensional parameter space we are able to achieve a very high probability in statistical identification of the evolved, e.g. old, and evolving, e.g. young, galaxy groups using purely optical observations. These parameters can be obtained from the optical observations of galaxies alone which are now available from various imaging/spectroscopic surveys. 

We note that while the conventional definition of the fossil groups, based on luminosity gap, can result in a contaminated sample of old groups, other probes such as the relaxed X-ray morphology and the co-centring of the X-ray emission and the dominant galaxy in fossil increase the probability of these galaxy groups being old. Observationally, most of the well studied fossil groups \citep{Khosroshahi2004a,Sun2004,Khosroshahi2006,Khosroshahi2007} meet these criteria. 

We show that old systems have more concentrated halos than young ones in various mass bins, in agreement with \citet{Khosroshahi2007} which showed that for a given optical luminosity, fossil groups have a more concentrated halo, compared to non-fossil groups. However, measuring the halo concentration is still subject to large observational uncertainties, which prevents us to use this parameter for a reliable halo age dating in combination with other observables, routinely available from optical observations of galaxies.

We note that the there may some differences in the galaxy properties assigned to sub halos in different semi-analytic models \citep{Bower2006, DeLucia2007, Guo11} but this can only have a minor impact on the the values of the boundaries between the subregions in the parameter space introduced here. 

\section*{Acknowledgments}
The Millennium  Simulation database used in this paper and the web application providing online access to them were constructed as part of the activities of the German Astrophysical Virtual Observatory. We thankfully acknowledge Gerard Lemson for facilitating the  access to the data.

\bsp

\label{lastpage}

\end{document}